\documentclass[12pt]{article}
\usepackage[cp1251]{inputenc}
\usepackage[T2A]{fontenc}
\usepackage[dvips]{graphicx}
\newcommand{\bx}{{\bf x}}
\newcommand{\bv}{{\bf v}}
\newcommand{\bF}{{\bf F}}
\newcommand{\br}{{\bf r}}
\newcommand{\bV}{{\bf V}}
\newcommand{\bg}{{\bf g}}

\newcommand{\cR}{{\cal R}}

\newcommand{\rf}[1]{(\ref{#1})}

\newcommand{\beq}[1]{\begin{equation} \label{#1}}
\newcommand{\eeq}{\end{equation}}
\newcommand{\beqa}[1]{\begin{eqnarray} \label{#1}}
\newcommand{\eeqa}{\end{eqnarray}}
\newcommand{\barr}[1]{\begin{array}{#1}}
\newcommand{\earr}{\end{array}}
\newcommand{\beqn}{\begin{eqnarray*}}
\newcommand{\eeqn}{\end{eqnarray*}}
\newcommand{\prt}{\partial}
\newcommand{\bnab}{\nabla}

\newcommand{\veps}{\varepsilon}

\newcommand{\ga}{\alpha}

\newcommand{\gl}{\lambda}

\newcommand{\gs}{\sigma}

\newcommand{\gk}{\kappa}

\newcommand{\gD}{\Delta}
\newcommand{\gd}{\delta}

\newcommand{\gT}{\Theta}

\newcommand{\rot}{{\rm rot}}
\renewcommand{\div}{{\rm div}}

\newcommand{\lln}{{\rm ln}}
\newcommand{\const}{{\rm const}}

\title{The dust disk  dynamics in week-nonlinear regime}

\author{Victor M. Zhuravlev, Alexander V. Patrushev }

\date{}

\begin{document}
\maketitle

\bigskip

\centerline{Ulyanovsk State University}

\bigskip
\centerline{zhuravl@sv.ulsu.ru}

\bigskip

{\small We investigate the problem of self-gravitating dust disk
dynamics in a static state taking into account nonlinear effects.
For this purpose Schr\"odinger-type equation including the mass
conservation law is used for the whole description of hydrodynamic
flows of self-gravitating dust. We have shown a purely
hydrodynamic mechanism of ring formation in the radial direction
by taking into account nonlinearity in the lowest order of
expansion parameter, which determines an order of magnitude of
flow. }
\bigskip
\section{Introduction}
\large

Self-gravitating systems are of great interest for investigation
in astrophysics because of their widespread appearance
\cite{Fr,Hu,Ber}. Such systems are difficult for analytical
analysis when we consider a lot of factors, and therefore the
standard way of analysis is in the terms of density perturbations.
As a result the problem may be entirely linearized, and it
facilitates analysis, but in this way one cannot entirely
investigate some specific effects caused by nonlinear dynamics of
such systems. In this paper the problem is also considered in
terms of perturbation theory, but it is possible to take into
account nonlinear effects of dynamics in the lowest order by means
of elimination of secular in time expansion terms.
Schr\"odinger-type equation\footnote{Of course we consider this
equation only as auxiliary one without Plank's constant and any
"quantum sense".} representing combined description of Euler fluid
dynamics together with the continuity equation is used for
determination of potential hydrodynamic flow. Dynamics of rather
outlying disk regions is discussed. These regions are situated far
from the central massive object and the disk surface. Internal
boundaries of these regions are predetermined by congruence
condition of the self-gravitating dust potential and the compact
body potential. The main aim of this paper is to consider a ring
structure formation due to nonlinear hydrodynamic flow of
self-gravitating dust. The internal region dynamics nearby the
central object and the disk surface must be considered as the
internal solution problem similarly to boundary layers problem.

\section{The usage of Schr\"odinger-type equation in Euler fluid dynamics}

    Let us consider  Schr\"odinger-type equation:
    \beq{Shred}
        {i\Psi_t+\ga(t)\gD \Psi - U(\bx,t)\Psi=0.}
    \eeq
    Here $\Psi$ is a dimensionless complex wave function, $i$ is an imaginary unit,
    $\ga=\ga(t)$ is a certain dimensionless real function of time $t$,
    $U(\bx,t)$ is a real potential-like energy function,
    $\gD$ is three-dimensional Laplacian.

    Besides the \rf{Shred} it is necessary to consider the complex conjugate equation:
    $$
        -i\Psi^*_t+\ga(t)\gD \Psi^* - U(\bx,t)\Psi^*=0.
    $$
    Multiplying the previous equation by $\Psi$ and \rf{Shred}
    by $\Psi^*$, and subtracting the second expression from the first one, we get
    \beq{ConsM}
        \frac{\prt}{\prt t}|\Psi|^2+\div\left(i\ga|\Psi|^2\bnab \lln
        \gT\right)=0,
    \eeq
    where $\gT=\Psi^*/\Psi$. Dividing \rf{Shred} and the conjugate
    equation by $\Psi$ and $\Psi^*$ correspondingly, and summing these expressions
    gives
    \beq{EqDyn}
     i\frac{\prt }{\prt t} \lln \gT +\ga \frac{\gD \Psi}{\Psi}+
     \ga \frac{\gD \Psi^*}{\Psi^*}-2U=0.
    \eeq
    \rf{ConsM} can be interpreted as differential conservation law for
     density $|\Psi|^2$ of fluid moving with  the velocity
     $\bv = i\ga\bnab \lln \gT$.

    For an arbitrary Euler flow $\bv$ we have the following identity:
    \beq{Eqvivl}
        \bv_t+(\bv,\bnab)\bv\equiv \frac{1}{2}\bnab
        |\bv|^2+[\bv\times\rot\bv]+\bv_t.
    \eeq
    Note that in our case the field $\bv$ is  potential:
    $\rot\bv=0$. For analysis of right-hand side of \rf{Eqvivl}
    one can use \rf{EqDyn}. As a result it follows the identity
    \beqn
        &&\bv_t+\frac{1}{2}\bnab |\bv|^2=\ga^2\bnab\left[-2\frac{\gD |\Psi|}{|\Psi|}\right]+2\ga\bnab U +\frac{\dot{\ga}}{\ga}\bv.
    \eeqn
    This result through identity \rf{Eqvivl} can be combined into
    the following Euler's equation for the flow $\bv$

    \beq{EqEuler}
        \bv_t+(\bv,\bnab)\bv = \ga\bnab\left[-2\ga\frac{\gD
        |\Psi|}{|\Psi|}+2 U\right]+\gk\bv,
    \eeq
    where $\gk=-\dot{\ga}/\ga=\gk(t)$ is a coefficient of a linear friction as it is called in fluid dynamics.
    Corresponding frictional force in the system $-\gk\bv$ can be regarded as a result of collisions
    between dust particles. In fact, this rather simplistic approach doesn't give an exact
    picture. Nevertheless, it allows to estimate the influence
    of a dissipation on a stationary dust distribution.
    Below we discuss the case $\gk=\gk_0=\const$. Dependence  $\gk$ on time can be interpreted as
    particles interaction to be changed in time, e. g. due to increasing particles in size .

    Consider force per unit mass in the momentum equation in details. One can see from \rf{EqEuler} the force is potential.
    In real fluid dynamics, however, right hand side in Euler equation in the presence of the field of a
    potential force is the following
    \beq{EqF}
        \bF=-\frac{1}{\rho}\bnab P -\bnab \phi,
    \eeq
    where $\rho$ is density of fluid or gas , $P$ is
    pressure, $\phi$ is potential of force per
    unit mass is considered  below to be Newtonian force.
    We investigate dust objects in this paper.
    And  the state equation for dust is well known as $p=0$.
    Therefore, force per unit mass \rf{EqF} consists of dust self-gravitation  and
    it may include gravitation force from a massive object, which is nearby the dust.
    Then we find the relation between Euler fluid dynamics and Schr\"odinger-type equation
    \beq{EqPres}
        -2\ga^2\frac{\gD
        |\Psi|}{|\Psi|}+2\ga U = -\phi.
    \eeq

    \section{ The hydrodynamic equations for self-gravitating dust }

    The problem considered in this paper can be formulated in the following way. We investigate
    dust objects having the equation of state $P=0$. And dust is
    in self-gravitation in terms of Euler fluid dynamics. So, we come to the system of equations for self-gravitating objects dynamics,
    which consists of
    Schr\"odinger-type equation \rf{Shred}, the concordance equation and
    the Poisson one, which is:

    \beq{EqPuas}
        \gD \phi = 4\pi G \Big(\rho_0 |\Psi|^2+\gs\gd(z)+M_0\gd(r)\Big),
    \eeq
    where $\rho_0$ - a characteristic density such that the function of density is
    $$
        \rho=\rho_0|\Psi|^2.
    $$

    The second z-direction $\gd$-like source item in right hand side of the equation for
    potential describes  a matter originally concentrated in thin
    disk with surface density $\gs=\gs(x,y,t)$, where $x,y$ are
    Cartesian coordinates at the disk surface.

    Now we shall make a nondimensionalization  as following:
    $$
         \tilde\br = \br/R_0,~~\tau=t/T_0,~~\Phi=\phi/\phi_0.
    $$
    And make a notion : $\ga(\tau)=\ga_0f(\tau)$, where
    $$
        f(\tau)=\exp\Big\{-\int\limits_0^\tau\gk(\tau')d\tau'\Big\}
    $$ is a dimensionless function of time.
    The system of equations then reads
        \beqa{EqShr}
        && i\Psi_\tau+\veps f(\tau)\tilde\gD \Psi - W\Psi=0,\\
        \label{EqCon}
        && -2\veps^2f^2(\tau)\frac{\tilde\gD
        |\Psi|}{|\Psi|}+2\veps f(\tau)W = - \frac{\phi_0
        T_0^2}{R_0^2}\Phi,\\
        \label{EqPuas}
        && \tilde\gD \Phi = \frac{4\pi G \rho_0 R_0^2}{\phi_0} |\Psi|^2,
    \eeqa
    where $\veps =T_0\ga_0/R_0^2$ - is the dimensionless characteristic parameter,
    estimating an order of magnitude of the dimensionless flow in the system:
    $$
        \bV = i\veps f(\tau)\frac{\prt}{\prt \tilde\br} \lln \gT,
    $$
    $W=U T_0$ is the non-dimensional collective potential.

    Below  we omit $\tilde{\ }$, implying the equations to be written
    in a dimensionless form.

     In this paper we are interested in cases such that
     $\veps << 1$ is a small parameter, i.e.  the velocity of flow is small
    and the system in a gravitation field is about equilibrium .

    Researching of \rf{EqShr}-\rf{EqPuas} shows that
    for the equations to describe non-trivial situation we must
    use the  conditions as follows:
    $$
        \frac{\phi_0 T_0^2}{R_0^2} = \veps,~~\frac{4\pi G \rho_0
        R_0^2}{\phi_0}=\mu = O(1),
    $$

    where $\mu$ is the first-order constant to $\veps$.
    From this it follows $\phi_0 =\veps R^2_0/T_0^2,~~\rho_0=\veps \mu /(4\pi G
    T_0^2)$, i.e. the dust density and the self gravitational
    potential are small and have the same order with respect to
    $\veps$.

    \section{ Approximate equations}

    Let us seek the solutions as power series in $\veps$:
    $$
        \Psi=\Psi_0+\sum\limits_{n=1}^\infty\veps^n\Psi_n,~~~
        \Phi=\Phi_0+\sum\limits_{n=1}^\infty\veps^n\Phi_n,~~~
        W=W_0+\sum\limits_{n=1}^\infty\veps^nW_n.
    $$
    Since small parameter $\veps$ in \rf{EqShr} and
    \rf{EqCon} is at the derivative of higher order, we can expect boundary layers in
    the system to  appear. They could exist in the center of the field and at the surface of the
    disk. This boundary layers are connected with nonlinear mode
    not with viscosity. Outside of this boundary layers, i.e. far from the field  center and the disk surface
    we use ordinary
    axial coordinate $z$.  Nearby the disk surface we must use coordinate $Z=z/\veps$.
    Note, $\gd$-like source in \rf{EqPuas} must be taken in account
    only in internal solution.

    For external region we have the system of equations by substituting the expansions
    in equations in two first orders :
    \beqa{Pow0}
        && i\Psi_{0,\tau} = W_0 \Psi_0;\qquad 2f(\tau)W_0=-\Phi_0;\qquad
        \gD \Phi_0 = \mu |\Psi_0|^2;\\
        \nonumber
        && i\Psi_{1,\tau}=W_0 \Psi_1+W_1 \Psi_0 - \gD \Psi_0,\\
        \label{Pow1}
        && -2f^2(\tau)\frac{\gD |\Psi_0|}{|\Psi_0|}+2f(\tau)U_1=\Phi_1,\qquad
        \gD\Phi_1 =
        \mu\Big(\Psi_0^*\Psi_1+\Psi_1^*\Psi_0\Big).
    \eeqa
    Suppose the flow and the gravitational field in lowest order are
    stationary; then we have solution in this order:
    $$
        \Psi_0=C_0(\br) \exp\left\{-i\int W_0(\br,\tau) d\tau\right\},~~
        W_0=-\frac{1}{2f(\tau)}\Phi_0,
    $$
    where function $\Phi_0$ is to be obtained from the Poisson equation:
    \beq{EqPhi0}
        \gD\Phi_0 = \mu |C_0|^2.
    \eeq

    After some simple manipulations, we arrive at the following solutions at the first order :
    \beqn
        &&\Psi_1=C_1(\br)e^{-i\chi(r,\tau)} -i e^{-i\chi(\br,\tau)}
        \int\limits_0^\tau
        \left[W_1(\br,\tau')C_0(\br)-
        e^{i\chi(\br,\tau')}f(\tau')\gD\Psi_0\right]d\tau',\\
        && W_1=-\frac{1}{2f(\tau)}\Phi_1+f(\tau)\frac{\gD |C_0|}{|C_0|}.
    \eeqn
    Here
    $$
        \chi(r,\tau)=\int W_0(\br,\tau) d\tau
    $$
    Substituting these expressions in the equation for $\Phi_1$
     \rf{Pow1}we get:
     \beqa{EqPhi1}
        &&\gD\Phi_1=\mu\Big(C_0C_1^*+C_1C_0^*\Big)+
        i\mu\Big[C_0^*\gD C_0-C_0^*\gD C_0\Big]Q(\tau)-
        \mu H(\tau)\bnab\left(|C_0|^2\bnab \Phi_0\right).
     \eeqa
     Here
     $$
        Q(\tau)=\int\limits_0^\tau f(\tau')d\tau',~~~~
        H(\tau)=\int\limits_0^\tau \int\limits_0^{\tau'}\frac{d\tau''}{f(\tau'')} f(\tau')d\tau',
     $$
     If $\gk=\gk_0=\const$ then
     $$
         Q(\tau)=\frac{1}{\gk_0}(1-e^{-\gk_0\tau}),~~
         H(\tau)=\tau\frac{1}{\gk_0}-\frac{1}{\gk^2_0}(1-e^{-\gk_0\tau}).
     $$
      One can see  that $Q(\tau)$ decays exponentially to a constant,
      whereas  $H(\tau)$  increases linearly with time, i.e. corresponding
      component is secular, and for stable solution it should become zero.
      Hence, we come to the following: taking into account \rf{EqPhi0}, functions
     $C_0(\br)$ and $W_0(\br)$   must obey the equations such that
     \beqa{EqC1}
        && \bnab\left(|C_0|^2\bnab \Phi_0\right)=0,\\
        \label{EqC0}
        && \gD \Phi_0 = \frac{\mu}{2} |C_0|^2.
     \eeqa
      Then equation \rf{EqPhi1} reduces to
     \beq{EqPhi2}
        \gD \Phi_1=\mu\Big(C_0C_1^*+C_1C_0^*\Big)
        +\frac{2\mu}{\gk_0}\div\Big[|C_0|^2\bnab \gT_0\Big]\left(1-e^{-\gk_0\tau}\right),
     \eeq
     where $C_1(\br)$ can be obtained from stationary condition at
     the next order of expansion
     , and $\gT_0=(i/2)\lln \Big(C^*_0/C_0\Big)$ -
     is still an arbitrary function. The solution for $W_1$ follows from the first equation in \rf{Pow1}.

     The interpretation of obtained equations follows from  the
     expression for the flow velocity in the first order

        $$
            \bV_1=\bv_1+\bv_2,~~~\bv_1= \veps\bnab \Phi_0, ~~~~
            \bv_2=2\veps f(\tau)\bnab \gT_0(\br) =2\veps e^{-\gk_0 \tau}\bnab \gT_0(\br).
        $$

     Right hand member in  \rf{EqPhi1} is determined by the source of
     mass, which is associated with the second flow in the system.
     This flow tends to a fixed space distribution as $\tau\to\infty$.
     Equation \rf{EqC1} is the law of conservation of mass for the  flow $\bv_1$.
     Suppose sources of mass do not exist; then the second  addend in right hand side of \rf{EqPhi1} becomes zero
     \beq{EqGT0}
         \div\Big[|C_0|^2\bnab \gT_0\Big]=0.
     \eeq
     If not, we must explicitly write down the source of
     mass by means of join of external and internal solution.
     The first flow is stationary and it is associated  in first
     order with stationary fall in the field $\Phi_0$. Existence of
     dissipation leads to fall of particles with the fixed velocity $\bv_1$
     instead of falling with the acceleration $\bg=-\bnab \Phi_0$.

      \section{Axial-symmetric solutions }
     Our aim is to consider models with axial symmetry.
     Take cylindrical polar coordinates $r,z,\varphi$ implying dependence functions of the system on $r$ and $z$
     $$
         \Phi_0(r,z)=u(r)h(z),~~\cR(r,z)=p(r)h(z).
     $$
     Thus one can find the following equations for $u(r),~p(r),~h(z)$:
     \beqn
        &&
        \frac{u''}{u}+\frac{1}{r}\frac{u'}{u}+\frac{h''}{h}=\frac{\mu}{2}\frac{p}{u},
        ~~~~~~\frac{u'}{u}\frac{p'}{p}+\frac{(h')^2}{h^2}+\frac{\mu}{2}\frac{p}{u}=0.
     \eeqn
     Separation of variables implies
     $$
        h(z)=h_0e^{-\gl z}.
     $$

     We should suppose dust density and potential to decay
     while moving off the disk in the line of $z\to+\infty$ as well
     as $z\to-\infty$. Thus if $\gl>0$, we have
     $h(z)=h_0e^{-\gl|z|}$ far from the disk surface $z=0$.

     So we obtain equations for $u$ and $p$
     \beqa{Equ}
        && u''+\frac{1}{r}u'+\gl^2u=\frac{\mu}{2}p ,\\
        \label{Eqp}
        &&
        \frac{u'}{u}\frac{p'}{p}+\gl^2+\frac{\mu}{2}\frac{p}{u}=0.
     \eeqa

     Let us seek the solution for $p$ as: {$p(r)=q(r)u'(r)$}.
     Substituting $p$ in above form  into equation \rf{Eqp} and using
     \rf{Equ} yield
     $$
         q'-\frac{1}{r}q+\mu q^2=0.
     $$
     This equation is easy to solve and general solution is
     $$
        q(r)=\frac{2}{\mu}\frac{r}{r^2+Q_0},
     $$
     here $Q_0$ - integral constant. Finally we get
     \beq{Eqpr}
        p(r)=\frac{2}{\mu}\frac{r}{r^2+Q_0}u'(r),
     \eeq
     where $u(r)$ now follows the equation
          \beq{EqWr1}
        u''+\frac{Q_0}{r(r^2+Q_0)}u'+\gl^2u=0.
     \eeq
     Function $p(r)$ follows, correspondingly, the equation
    \beq{Eqpr1}
        {p''+\frac{2r^2-Q_0}{r(r^2+Q_0)}p'+\gl^2p=0.}
     \eeq

     At small $Q$ or large $r$ the equation for $u(r)$ is similar to oscillation equation with wave number
       $\gl$ and, hence, $u(r)$ changes the sign quasi-periodically as well as its derivative.

    So we can establish behavior of dust
    distribution.
        The definition of $p(r)$ implies $p(r)>0$.
    The density becomes zero together with the gradient of potential
    according to \rf{Eqpr}. Thus it follows that disk is
    partitioned on rings which are separated by thin gaps.
    Analysis of equations for flow shows that dust from certain
    ring does not penetrate the bounders. The gradients of
    potential in adjacent rings, however, have unlike signs due
    to its quasi-periodic behavior. But if we require continuity of density and its derivative at boundary
    points, we immediately  obtain that density must attain a negative value
    along with  $u'$. Therefore, the derivative of density has
    discontinuity at boundaries of rings and in each ring with
    constant sign of $u'$ we should choose the sign and the magnitude of
    $Q$ to satisfy  requirement $p(r)>0$.

    Generally boundary conditions for calculation of ring
    parameters come to continuity of potential and its derivative
    at ring boundaries (second derivative is discontinuous). It
    follows from  equality of forces operating at disk boundaries.

    Let
    $r_i,~~i=1,2\ldots$ be boundary points in which $u'|_{r_i}=0$.
    Then equations for our model are:
    $$
        u_j''+\frac{Q_j}{r(r^2+Q_j)}u_j'+\gl_j^2 u_j=0,~~r\in[r_j,r_{j+1}]
    $$
    with boundary conditions
     \beqn
        && u'_j(r_j)=0,~~~u'_j(r_{j+1})=0,\\
        &&
        u_{j-1}(r_j)=u_{j}(r_j),~~u_{j+1}(r_{j+1})=u_{j}(r_{j+1}),\\
        &&  \pi\mu\int\limits_{r_{j}}^{r_{j+1}}\frac{
        r^2}{(r^2+Q_j)}u'_j(r)dr=\gs_j.
    \eeqn
    Here $\gs_j$ is a constant surface density in  $j$ ring.
       \begin{center}
        %\hspace{1cm}
        {
        \begin{center}
        \includegraphics[width=1\textwidth]{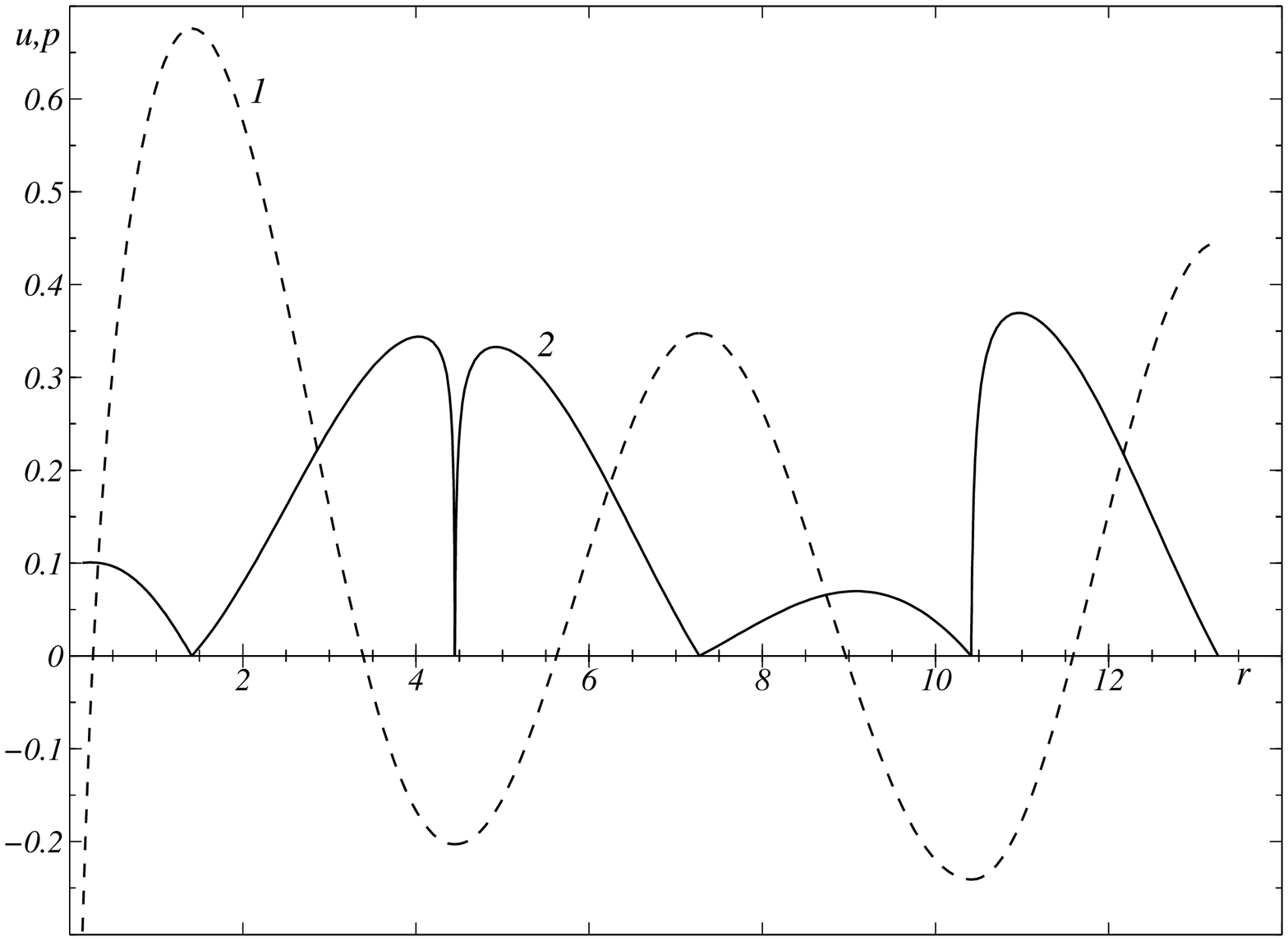}
        \end{center}
        }
        \bigskip
        {  Fig.1.  1 - potential $u(r)$ , 2 - density $p(r)$. Ring parameters :
        $\gl=1$, $\mu=2$,~$Q_0=10,~Q_1=-19.85,~Q_2=-19.7,~Q_3=-155,~Q_4=-108$. }
        \end{center}

    Figure 1 illustrates the different possible solutions satisfying  the boundary
    conditions.

    \section{Discussion}

        We have investigated the formation of rings
     with too narrow gaps between them  as $t\to\infty$.
     And the width of gaps are considerably smaller
     then disk rings one.
     The masses of rings and distribution of density are
     determined by the coefficient $Q$, that can be unique for each
     ring.
     Obtained solutions can be associated with  dust distribution
     in real disk systems. However, to describe  systems like
     internal Saturn's rings we must use other methods because
     our approach is not suitable for internal regions.
     Radial orbital flow induced by the central mass is the main component of dynamics
     for internal rings. Nearby the planet the radial velocity is
     great but we assume the mechanism of ring formation to be
     the same and it can be modified by corrections connected with
     the main orbital flow. And it will be the object of another paper.


\begin{thebibliography}{99}
     \bibitem{Fr}Fridman, A. M. and Polyachenko, V. L. (1984) Physics of Gravitating Systems, Vols 1
      and 2. Springer, New York.
     \bibitem{Hu}Hubert Klahr and D. N. C. Lin IdentifiersDust Distribution in Gas Disks. II. Self-induced Ring Formation through a Clumping Instability
       ApJ,2005, 632,1113.
     \bibitem{Ber}James R.Graham, Astrophysical Gas Dynamics. http://astron.berkeley.edu/~jrg/ay202/
    \end{thebibliography}
\end{document}